# Layered Bimetal Nanoporous Platforms for SERS Sensing

*Yanqiu Zou[1,±], Anastasiia Sapunova[2,3,±], Tommaso Giovannini[4], Chen Wang[5], Huaizhou Jin[6], Vincenzo Caligiuri[7], Andrea Schirato[8,10], Luca Bursi[9], Alessandro Alabastri[10], Shukun Weng[2,3], Ali Douaki[2,11], German Lanzavecchia[2,11], Ivan Marri[11], Roman Krahne[2], Nicolò Maccaferri[12,13], Zhenrong Zheng[1,\*], Shangzhong Jin[5\*] and Denis Garoli[\*2,5,11]*

[1.] State Key Laboratory of Modern Optical Instrumentation, College of Optical Science and Engineering, Zhejiang University, Hangzhou 310027, China.

[2.] Istituto Italiano di Tecnologia, Via Morego 30, 16163 Genova, Italy.

[3.] Università degli Studi di Milano-Bicocca, Piazza dell'Ateneo Nuovo 1, 20126, Milano, Italy

[4.] Department of Physics, University of Rome Tor Vergata, and INFN, Via della Ricerca Scientifica 1, I-00133, Rome, Italy

[5.] College of Optical and Electronic Technology, China Jiliang University, Hangzhou 310018, China

[6.] Key Laboratory of Quantum Precision Measurement, College of Physics, Zhejiang University of Technology, Hangzhou, China.

[7.] Dipartimento di Fisica, Università della Calabria, via P. Bucci 33b, 87036 Rende, CS, Italy

[8.] Dipartimento di Fisica, Politecnico di Milano, Piazza Leonardo da Vinci 32, I-20133 Milano, Italy

[9.] Department of Physical, Computer, and Mathematical Sciences (FIM), Università degli Studi di Modena e Reggio Emilia, and Nanoscience Institute, CNR-NANO, via Campi 213/a, 41125, Modena, Italy

[10.] Department of Physics and Astronomy, Rice University, Houston, Texas 77005, United States

[11.] Università degli Studi di Modena e Reggio Emilia, Via Amendola 2, 42122, Reggio Emilia (Italy)

[12.] Department of Physics, Umeå University, Linnaeus väg 24, 901 87 Umeå, Sweden

[13.] Umeå Centre for Microbial Research, Umeå University, 901 87 Umeå, Sweden

±Equally contribution

* Corresponding author: Denis Garoli, denis.garoli@unimore.it; Zhenrong Zheng, zzr@zju.edu.cn; Shangzhong Jin, Jinsz@cjlu.edu.cn

**Abstract**

Nanoporous metals are extensively investigated as platforms for applications in plasmonics. They present high surface areas and strong local electric fields that can be tuned at different energies, playing with the choice of the metals and the morphology of the porous layers. Until recently, research in the field of plasmonics has primarily focused on porous metals composed of a single element, with limited attention given to the impact of alloy composition. The investigation of bi-metallic systems has only just begun to emerge in the literature. In particular, combining two or more different plasmonic metals, it could be possible to explore the interactions between two metals excited at specific energies. This involves plasmonic coupling, electron transfer, band hybridization at the interface, electromagnetic field interactions, and possibly thermal and electronic energy transfer depending on separation, size, and materials involved. The analysis of bi-metal systems can also be interesting in biomolecule detection, such as in the case of Surface Enhanced Raman Scattering (SERS). Here we report, for the first time, a detailed study (comprising morphological analyses, numerical modelling, and optical spectroscopies) on bi-metal nanoporous platforms prepared with a dry-synthesis method enabling the easy and controllable fabrication of bilayers combining different metals such as Au, Ag, and Cu.

**Materials and Methods:**

*Samples fabrication*

The samples preparation is based on the original methods proposed by Kwon et al.[1]. In brief, poly (methyl methacrylate) (PMMA) was spin-coated on a Si substrate at 4000 rpm for 2 min. Each metal (>99.99 % purity) was subsequently deposited by electron-beam evaporation onto the PMMA thin film at room temperature with an oblique incidence angle of 80°, a deposition rate of 0.1 nm/s, and a target thickness of about 12 nm. The deposited Au film was plasma treated in $O_2$ at a power of 200W until the entire PMMA layer was removed. In the case of Ag and Cu films, in order to avoid fast oxidation, the plasma treatment was performed in $N_2$ at 200W until the PMMA layer was completely removed (as previously demonstrated, this ensures that the oxidation is kept to a very low level[2]). The bi-layer structure was obtained by repeating the preparation using the first porous layer as substrate for the deposition of the second porous layer on top.

*Spectroscopic measurements*

Raman spectroscopy measurements were performed using a HORIBA LabRAM HR Evolution Raman spectrometer (HORIBA Jobin Yvon, Kyoto, Japan) with a 50X long-focal-length objective (NA = 0.75), a 100μm-hole aperture, and a 600 gr/mm grating. SERS spectra of Rhodamine 6G (R6G) were acquired under different conditions to characterize system performance: 100 nM R6G with laser power attenuated to 10% at 532 nm excitation; 1 μM R6G with 5% laser power at 633 nm; and 0.1 mM R6G with full laser power (no attenuation) at 785 nm. For each substrate, a Raman map was collected over a 20 × 20 μm² area with a step size of 5 μm, resulting in 25 individual spectra. The final spectrum for each substrate was generated by averaging these 25 spectra after pre-processing. Spectra of the protein ADAMTS3 were obtained by averaging five randomly collected spectra, using a 5% ND filter and 60s acquisition time for both 532nm and 633nm excitations. For 785nm excitation, a 100% ND filter (no attenuation) and 120s acquisition time were used.

*Data Analysis*

The raw spectra were baseline-corrected and smoothed using HORIBA LabSpec6 software. Subsequently, all data averaging, normalization, spectral deconvolution, peak fitting, and graphical plotting were performed using OriginPro 2025b.

*Numerical modelling*

To explore the plasmonic properties of porous NPMs, we performed a numerical investigation of the electromagnetic response using a finite element method (FEM) commercial software, employing COMSOL Multiphysics. In particular, following a procedure reported in detail in our recently reported works[1,2] nanometric pores and irregularities of NPMs have been numerically reproduced from scanning electron microscopy (SEM) images of the experimental samples. However, continuum modeling, such as FEM, might fail in describing the response properties of bimetallic nanostructured materials[3–5]. Therefore, to validate the FEM procedure, the optical response of NPMs has also been calculated using a fully atomistic approach, named frequency-dependent fluctuating charges and dipoles (ωFQFμ)[5–7]. This method provides a reliable description of noble metal nanoparticles, comparable to *ab initio*[5] but at significantly lower computational cost, and can thus handle systems composed of thousands of atoms[8–11]. Remarkably, it can also be applied to bimetallic systems and, in the specific case of Au/Ag nanoparticles, yields results in almost perfect agreement with experiment[12]. For cross-validation of the COMSOL results, we have adopted an idealized picture in which the same porous structure is generated and then considered either as Ag or Au, which are characterized by the same lattice constant (4.08 Å). Such structures were created by introducing randomly generated nanometric pores, thereby generating realistic surface roughness and porosity. In particular, we modeled a $7 \times 7 \times 4$ (nm$^3$) structure, composed of 6584 atoms, a size unfeasible for *ab initio* atomistic models (see SI for a graphical representation). The same nanoporous geometry was then used to simulate Ag/Ag, Au/Au, and mixed Au/Ag bilayers, allowing for a direct comparison of their optical response and a detailed analysis of the different compositional effects. All ωFQFμ calculations are performed using the Ag/Au parameters reported in Ref. 12, with the open-source plasmonX software[6].

*Dielectric Permittivity and Ellipsometry.*
Spectroscopic ellipsometry was employed to determine the dielectric permittivity of nanoporous Ag/Ag, Ag/Au, and Ag/Cu bilayers. The measurements provided both the real and imaginary components of the dielectric function over the visible range. The experimental data were fitted using a model composed of Lorentzian and Gaussian oscillators to capture the main features of each alloy's optical response. The resulting dielectric permittivities accurately reproduce the experimental ellipsometric parameters Ψ and Δ. Complete fitting details and oscillator parameters are reported in the Supporting Information (Table S1).

**Introduction**

Over the last decades, research on nanoporous metals (NPMs) has grown steadily, driven by their broad range of applications in sensing[13–15], catalysys[16,17], photonics[18,19] and biomedicine[20]. The main characteristic of NPMs is their high surface-to-volume ratio that, combined with unique chemical reactivity and mass transport properties, makes this family of nanostructured materials highly interesting for advanced research. NPMs can be prepared following different strategies such as template-assisted physical vapor deposition, electrochemical processes, and chemical dealloying[18,21]. The latter is probably the most used method to prepare NPMs through selective dissolution of the more chemically active component from precursor alloys. Dealloying methods have been applied to prepare NPMs as gold[13,18,22], copper[23,24], silver[25,26,] and aluminum[27,28]. A specific aspect of dealloying is the presence of residual less noble metal at the end of the process. This effect is difficult to avoid and leads to resulting nanoporous structures with varying alloy compositions[29,30]. Interestingly, this residual "contamination" in the final porous film, long considered as a limitation in the preparation of pure porous structures, has recently been shown to enhance plasmonic properties of nanoporous gold containing residual amount of the less noble metal due to modulation of its permittivity by the alloying components.[31–33] The combination of two or more metals in alloys is a strategy to provide a way to tune the electronic state through charge transfer between the metals, making it possible to alter the chemical interaction between molecules and the surface of the alloy. In particular, the adsorption behavior of molecules having either high electronegativity (e.g., oxygen and halogens), strong polarity (e.g., carbonyl and nitro groups), partial charge (e.g., carboxylate and hydride-like hydrogen), or high electron density (e.g., alkenes and arenes) is affected by the alloy composition. For instance, palladium-silver alloy promoted the adsorption of negatively charged molecules (carboxylate intermediate) due to the high electronegativity of silver. The alteration of the electronic state by alloying on the molecular detection capability has been recently studied by La et al.[33] In particular, a SERS enhancement of up to 6 orders of magnitude for nanoporous gold containing silver was observed and attributed to the narrowing of the electronic structure and its alignment close to the Fermi level. These results suggest that the electronic state modulation of the metal structure can affect molecule-surface interactions.

Here, we decided to explore bi-metallic nanoporous systems from a different point of view. Instead of using chemical dealloying, we used a recently reported dry-synthesis method to prepare highly pure layers of plasmonic metals such as Au, Ag, and Cu[1–3]. The peculiarity of this approach lies in the possibility to produce stacked nanoporous layers, either composed of the same metal or of different metals deposited sequentially on top of one another, as demonstrated previously in refs. [2,3] At the best of our knowledge, this is the first example of perfectly stacked nanoporous metal layers, and it represents a valuable platform to investigate how the use of two different metals can impact the performance in terms of SERS enhancements at different wavelengths across the visible spectral range. Experimental results and numerical simulations will be presented to shed light on the behavior of these novel plasmonic nanoporous platforms.

**Results and discussion**

Figure 1 illustrates the planar morphologies of the samples used for these experiments. While the first row (Fig. 1(a-c)) reports the single NPM films, respectively Au, Ag, and Cu, the other micrographs show the obtained morphologies for the bi-metallic layered films. As can be seen, the single metal films present a nanoporous structure with fully connected ligaments of diameters of a few tens of nm. In the case of bi-metallic layered films, the morphologies vary significantly among the samples. In all the bi-metallic samples (Fig. 1(d-i)), it is possible to observe the two layers of metals stacked one on the other, but not mixed, with a gap between the 2 metallic nanostructures that can only be due to a potential small oxide layer surrounding the Ag and Cu nanoporous structures.

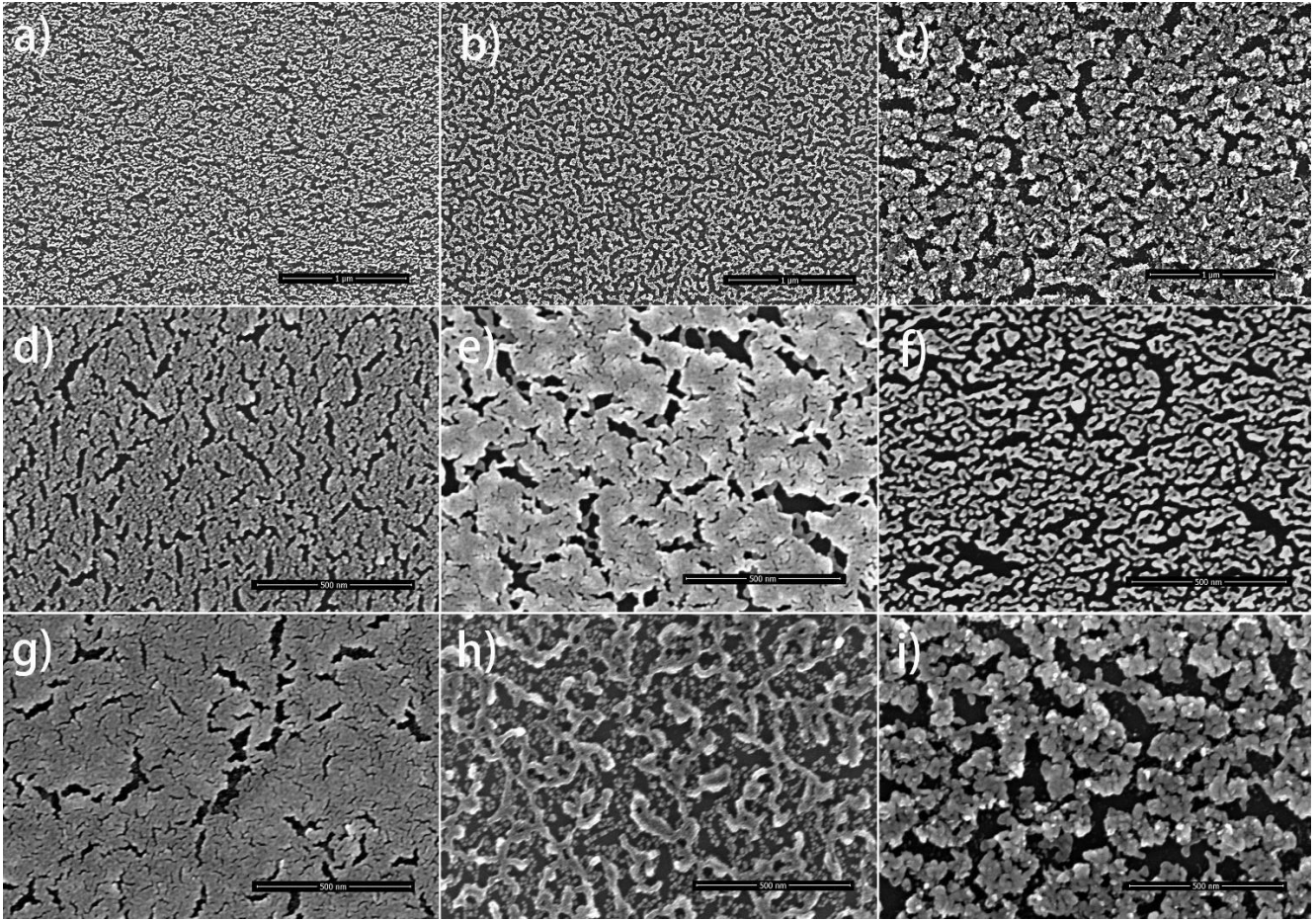

***Figure 1:*** *(a-e) SEM micrographs of the investigated samples; porous layers of (a) Au; (b) Ag; (c) Cu; (d) Au/Ag; (e) Ag/Au; (f) Au/Cu; (g) Cu/Au; (h) Ag/Cu; (i) Cu/Ag.*

Following the same approach used in our recent works[3,4], we performed numerical studies of the electromagnetic response of the porous bi-layer system. This enables us to obtain an overview of the electromagnetic (EM) field distributions at the wavelengths of excitation used in the experiments. The nanometric structural features of the layers were reproduced directly from the SEM images of the experimental samples in the finite element method (FEM)-based model, developed using COMSOL Multiphysics. Specifically, the top-view SEM micrographs of the Au, Ag, and Cu films used in the experiments were imported into the FEM solver and extruded along the thickness direction of the layer to build the 3D numerical geometry. As previously shown[3], this method enabled us to numerically account for the actual discontinuities and profile of the materials down to the nanoscale. Figures 2, 3, 4, and 5 report an estimation of the EM field enhancement distribution along the surfaces of the two (the top and bottom) porous layers in the different cases, taking into consideration three wavelengths of excitation (550, 630, and 780 nm).

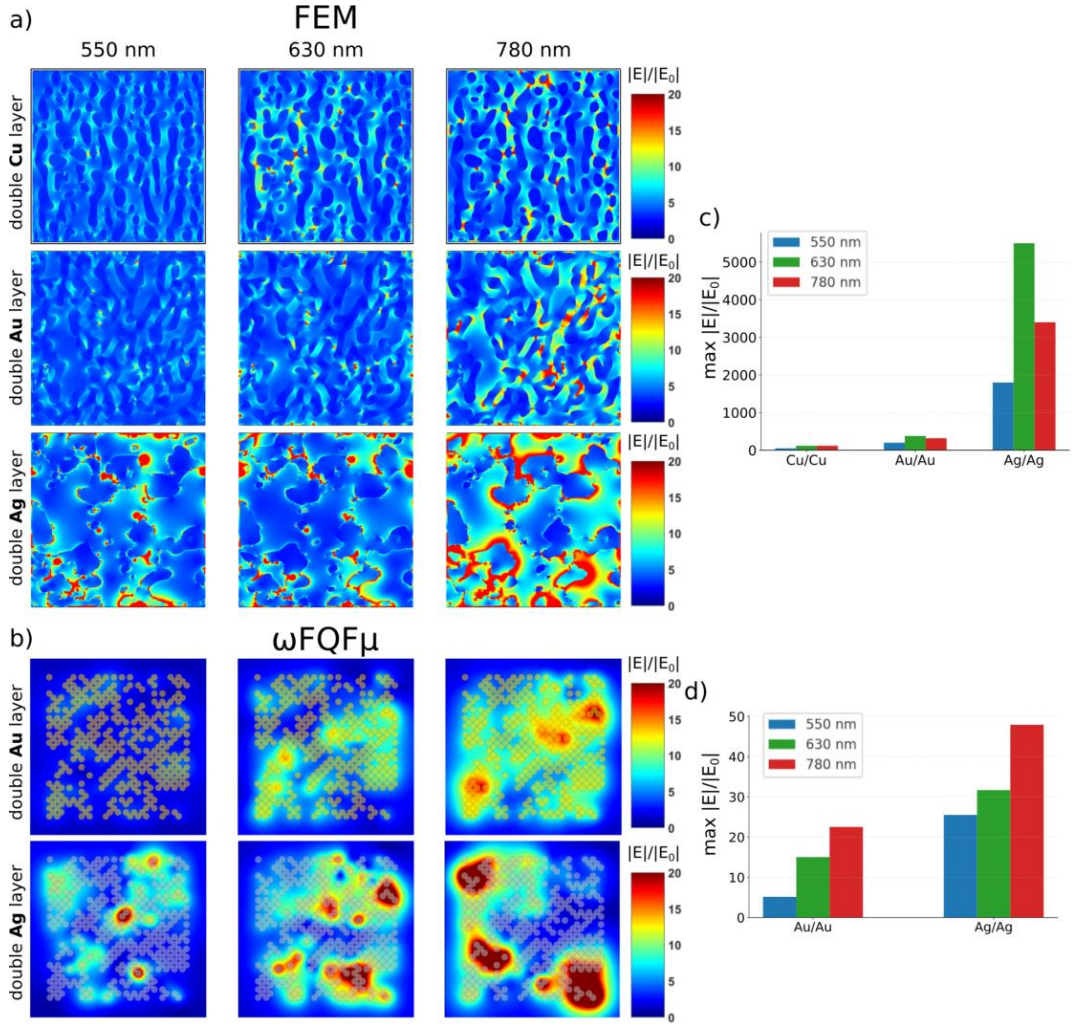

*Figure 2: (a,b) EM field enhancement distributions for (a) homometallic nanoporous bilayer structures composed of Cu (top), Au (middle), and Ag (bottom), calculated with FEM at excitation wavelengths of 550 nm, 630 nm, and 780 nm (from left to right); (b) homometallic nanoporous bilayers composed of Au (top) and Ag (bottom), obtained using the ωFQFμ approach at the same excitation wavelengths. All*

*maps are normalized to the same color scale to highlight spatial regions of strongest field enhancement localization. (c,d) Maximum field enhancement values extracted from FEM (c) and ωFQFμ (d) simulations for each structure.*

Figure 2 shows the EM field distributions at wavelengths of 550, 630, and 780 nm for bi-layer nanoporous structures made of a single metal: Cu (a, top), Au (a, middle), and Ag (a, bottom). For the specific case of Au and Ag, the EM distributions are also computed using the ωFQFμ method, for comparison (Fig. 2b). In all cases, the EM enhancement profiles were evaluated on the surface of the bi-layer structure. The color scale indicates a normalized intensity of the EM field, calculated as the ratio of the electric field in the presence of nanoporous metals to the incident field. All field maps are normalized to the same values, pointing out the regions where the EM field is highly localized.

Nevertheless, at the FEM level (Fig. 2a), each homometallic sample is characterized by its own geometry and plasmonic properties; therefore, the diagram of the maximum intensities of the EM field is shown for each case (Fig. 2c). The strongest enhancement across all studied wavelengths is observed for Ag at 630 nm, whereas Cu shows the weakest response, with almost negligible activity at 550 nm. For Cu and Au, the localized enhanced fields are mainly observed in the pore gaps, whereas in nanoporous Ag, there is a large number of "hot spots" strongly affecting the distribution of the field. Similar results are also obtained at the ωFQFμ level (see Fig. 2b) for Ag and Au NPMs. In this case, the electric field shows a different enhancement distribution depending on the metal and the laser frequency, reflecting a different induced density on the surface. In particular, as for FEM results, the response of Ag NPM is significantly larger than that of Au NPM, and also more hot-spots are observed. Nevertheless, the trends of the maximum electric field enhancements do not correlate exactly with those computed by using FEM (see Fig. 2c-d): in fact, in both Ag and Au NPMs, the maximum enhanced field is reported at 780 nm. This can, however, be due to the diverse plasmonic response provided by the two models, and by the small model structure considered in ωFQFμ calculations.

Understanding the optical response of homometallic nanoporous structures at different excitation wavelengths is crucial for a deeper comprehension of the processes occurring across the layers in bi-metallic configurations and for disentangling the individual contribution of each metal. Accordingly, new plasmonic effects emerge in Au/Ag, Ag/Au, Au/Cu, Cu/Au, Ag/Cu, and Cu/Ag NPM systems as a result of interlayer coupling and interactions between the constituent metals. Figure 3 depicts the EM field enhancement distributions for Au/Ag and Ag/Au NPMs (a-b) and field enhancement maxima (c-d) at different wavelengths computed by using FEM (a-c) and at the ωFQFμ level (b-d). At both levels, in the Au/Ag bilayer, the silver layer (Figure 3a), which is placed on top of Au displays significantly reduced EM field intensity at all studied wavelengths relative to the homometallic Ag layer (Figure 2a-b). Nevertheless, the Ag layer exhibits stronger plasmonic activity, despite the less intense localization of hot spots caused by the electrodynamical interaction between two porous layers consisting of two different plasmonic materials. Interestingly, at the FEM level, illumination of the Au/Ag NPM at 780 nm produces a field enhancement comparable to that at 630 nm (Fig. 3c, top), which can likely be associated with the interaction between plasmonic modes of silver and gold. This is also confirmed by the field maps computed at the ωFQFμ level and by the maximum field enhancements, which are similar at all considered frequencies (Fig. 3d, top). The same interaction induces a stronger enhancement for Au layer (Figure 3c-d) at 550 nm than in the homometallic layer (Figure 2c-d) and rises with increasing wavelength, a trend that is perfectly reproduced qualitatively by both methods.

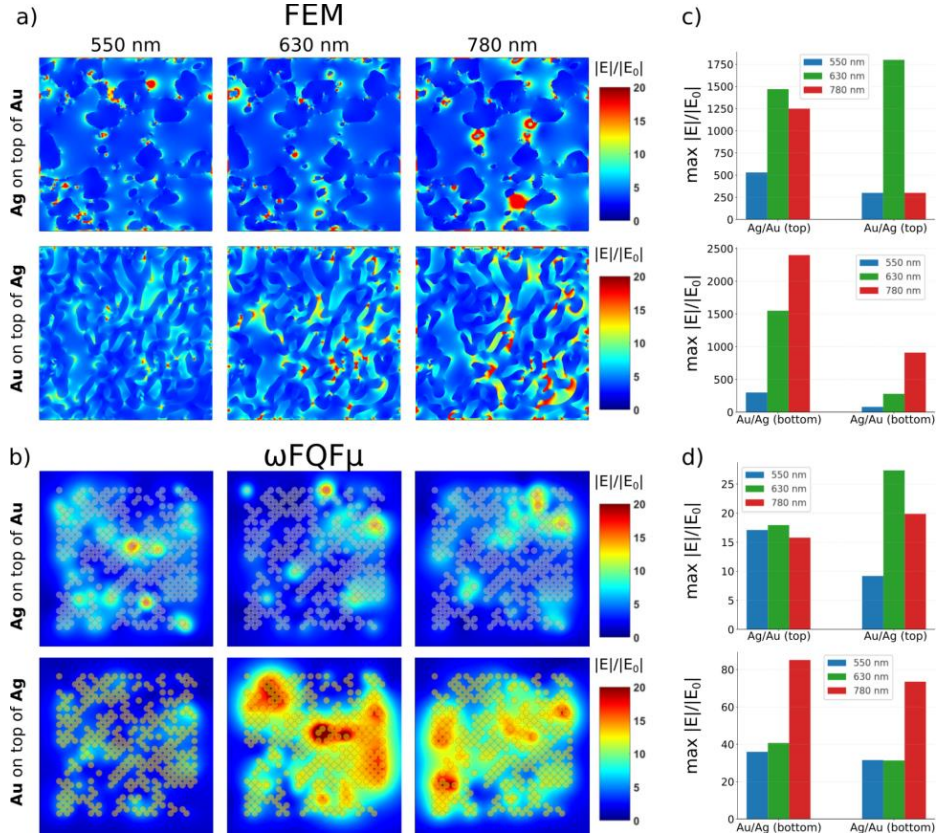

***Figure 3:*** *(a-b) The enhanced EM field distribution of Au and Ag layers for Au/Ag (top) and Ag/Au (bottom) NPMs calculated at the FEM (a) and ωFQFμ (b) levels of theory. (c-d) FEM c) and ωFQFμ (d) field enhancement maxima for each structure on top and bottom sides.*

In contrast, when the Au layer is on top of Ag (Figure 3a-b, bottom), the plasmonic response remains weaker at 550 and 780 nm, but increases significantly at 630 nm because of interlayer coupling. This remarkable result, reproduced by both numerical methods (see Fig. 3c-d, top), indicates that the stacking sequence strongly affects the resonance conditions. As a result, at 630 nm, silver has a weaker plasmonic response than in the homometallic Ag case. Between two systems, Au/Ag and Ag/Au configurations at 550 nm, the enhancement is better in the first configuration when the Ag layer is on top, consistent with the superior visible range plasmonic activity of Ag and its more efficient near-field localization. The wavelength-dependent trends of the enhanced field maxima predicted by the two numerical methods are qualitatively identical, validating the FEM-based approach for the subsequent analyses.

Although Au and Cu do not show the strong plasmonic response of Ag (Figure 2), their coupling still produces noteworthy effects. Figure 4 compares the EM field enhancement distributions and maxima for Au/Cu and Cu/Au bilayer configurations. In the case of Cu as upper layer (Figure 4 (a-c)), low plasmonic enhancement of Cu is compensated by coupling with underlying Au, especially at longer wavelengths (780 nm). When Au is on top of Cu (Figure 4 (g-i)), instead, the enhancement is mainly determined by the Au layer itself, while Cu acts as a substrate preventing additional EM field enhancement, possibly because of significant losses in the Cu layer. Interestingly, the field EM enhancement intensities of Cu are mostly the same in combination with Ag (Figure 5).

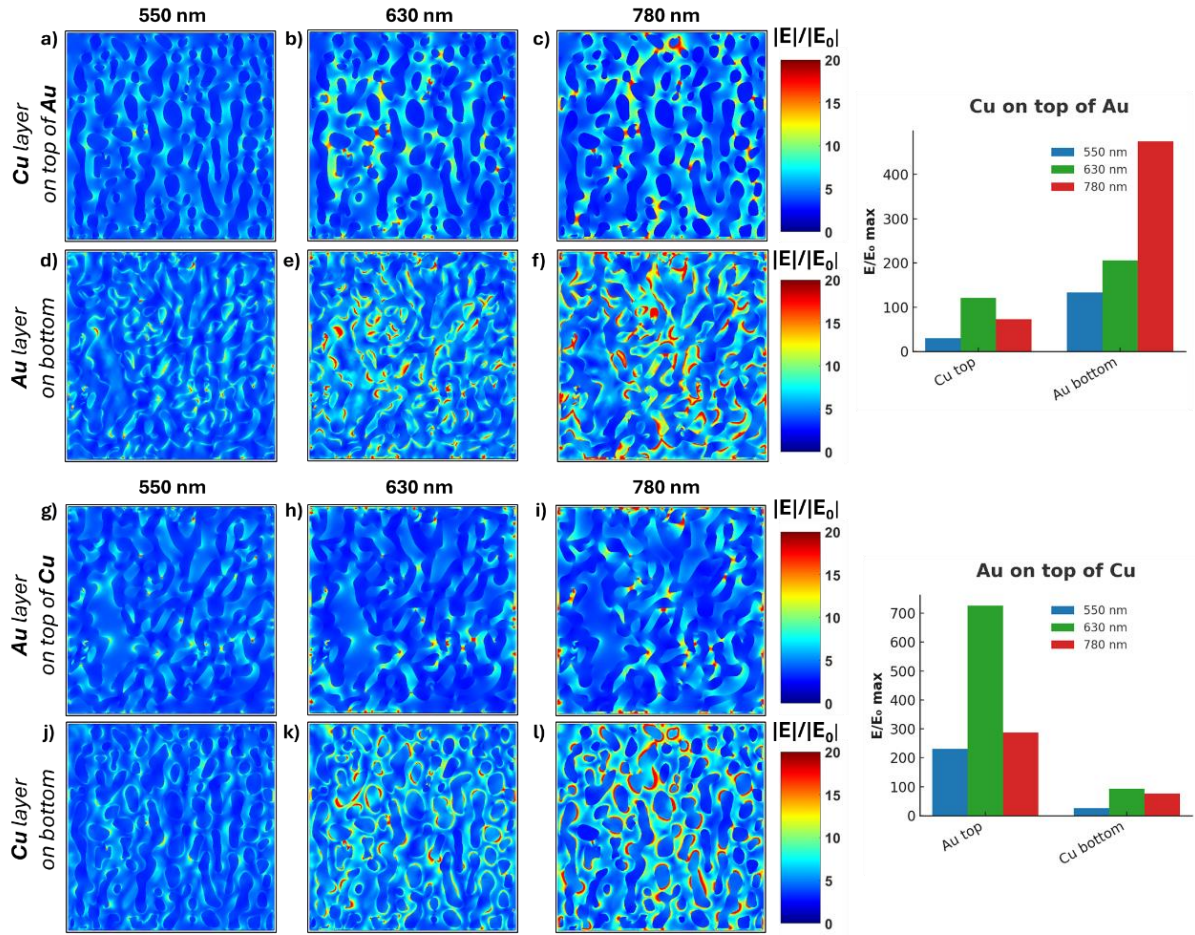

***Figure 4:*** *The enhanced EM field distribution of Au and Cu layers for (a-f) Au/Cu and (g-l) Cu/Au NPMs calculated as a ratio of the electric field in the presence of nanoporous metals to the incident field for different wavelengths.*

In bimetallic structures Ag/Cu (Fig. 5 (a-f)) and Cu/Ag (Fig. 5(g-l)), the EM field enhancement is dominated by silver, while copper mainly shows negligible activity, as can be seen from a comparison between the EM field enhancement maxima of these configurations and those of the homometallic structure composed of Ag (Fig. 2a, bottom). For Cu located on top of Ag, the lower Ag layer still supports a strong EM field localization, though slightly reduced due to copper-induced losses. In contrast, when Ag is deposited on top of Cu, the EM field enhancement is intensively concentrated in the upper Ag layer, but the overall response, as mentioned earlier, is reduced compared to Ag alone due to the lossy copper layer. These results confirm that the stacking order determines whether the Cu layer suppresses the Ag layer's resonances or favors the interlayer EM interactions.

These conclusions are also supported by ellipsometric investigations that allowed us to retrieve the dielectric permittivities of the bi-metallic combinations (see Supporting Information). In particular, homometallic bi-layer Ag/Ag exhibits a pronounced metallic response with a negative real permittivity throughout the visible range, confirming its validity as a plasmonic material and efficient field confinement. Ag/Au shows broader dispersion and larger optical losses, reflecting its intrinsically damped plasmonic character. Conversely, Ag/Cu displays a largely suppressed plasmonic response, with the real part of $\varepsilon$ only weakly negative and significant absorption across the visible spectrum.

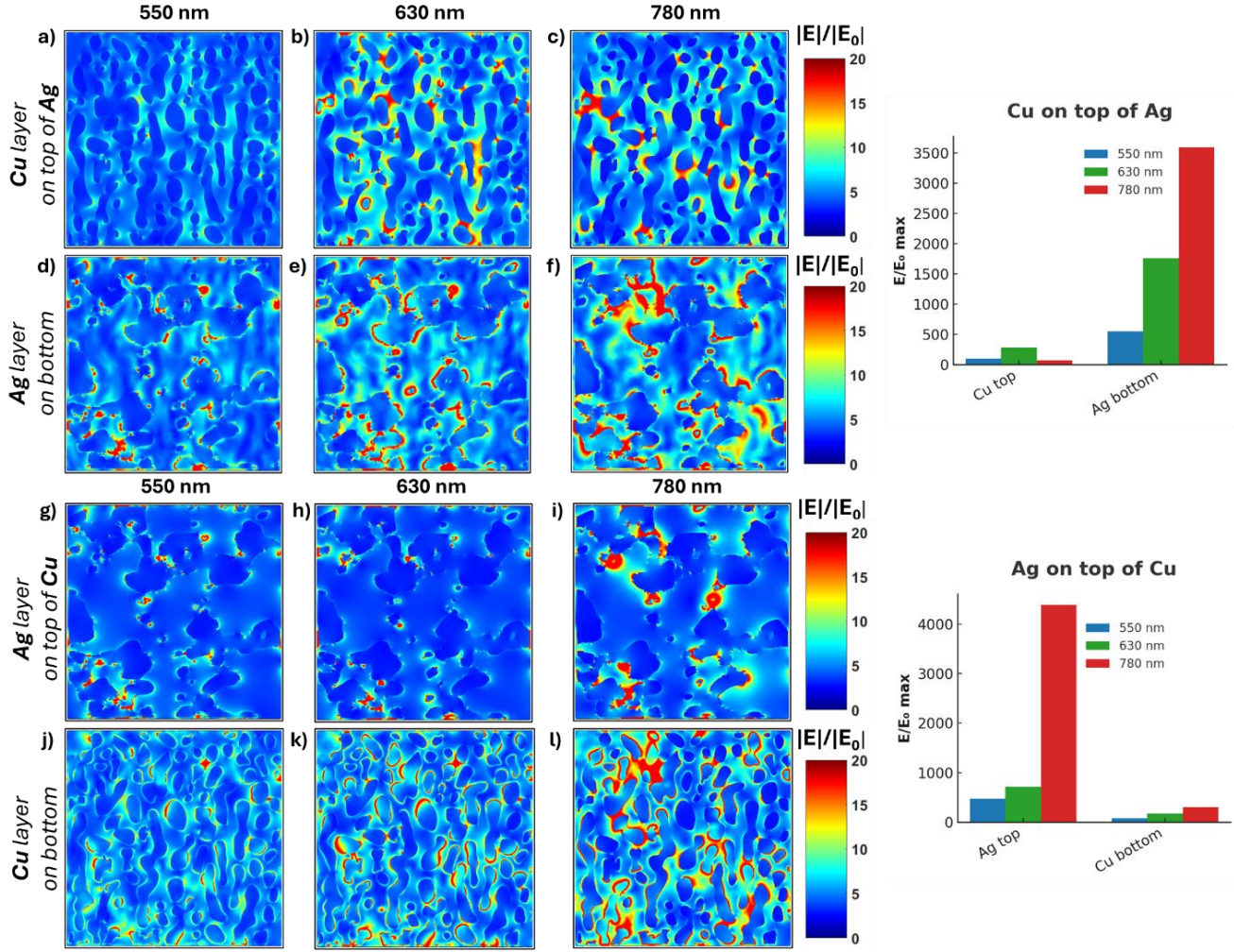

***Figure 5:*** *The enhanced EM field distribution of Ag and Cu layers for (a-f) Ag/Cu and (g-l) Cu/Ag NPMs calculated as a ratio of the electric field in the presence of nanoporous metals to the incident field for different wavelengths.*

The numerical results provide a framework for understanding and analyzing data during SERS experiments on the investigated bi-metallic NPMs. As predicted by our calculations, the efficiency of EM field enhancement and distribution in bi-layer systems is determined not only by the type of metal, but also by the order of the layers and the coupling efficiency between them. In homometallic structures, the enhancement is defined by plasmonic properties of each metal; moreover, the Fermi level remains the same. At the excitation energies explored here, in Au–Ag systems, silver dominates in terms of enhancement intensity, whereas the enhancement of gold is explained by the coupling of plasmonic modes. In the case of Ag–Cu, the interlayer coupling is weak, and the enhancement is determined mainly by silver. For Au–Cu, gold dominates in intensity. Additionally, the observed enhancement is still higher than in homometallic structures but remains significantly lower compared to configurations containing Ag.

Starting from the numerical prediction of the EM field distributions in our samples, the experimental performance in terms of SERS enhancement of the NPM films was examined under conditions as close as possible to the simulated ones. In particular, three excitation wavelengths very close to those analyzed in the simulations, i.e., 532 nm, 633 nm, and 785 nm, were employed experimentally, and Rhodamine 6G (R6G) was used as the molecular probe. The average SERS spectra of R6G, measured across the series of NPMs configurations with distinct bi-layer stacking sequences (e.g., Au/Cu, Ag/Au), are compared in Fig. 6 (a-c) to evaluate their corresponding SERS enhancement performances. Fig. 6 (d-f) presents the intensities of five characteristic R6G Raman bands at 620, 1280, 1360, 1509, and 1649 cm$^{-1}$ for the top three NPMs configurations, which were selected based on their highest overall SERS intensities under identical measurement conditions, as marked by asterisks in Figures 6 (a-c). To amid potential peak shifts, each intensity value was determined by taking the maximum within a ±10 cm$^{-1}$ window around the nominal wavenumber. The SERS spectra and corresponding peak intensities of R6G in Fig. 6 vary markedly across the various substrates and excitation wavelengths. Specifically, under 532 nm excitation, the average spectra of 100 nM R6G show that samples containing Ag as the bottom layer, combined with either Ag or Cu layers on top, yield the highest signal enhancement. The Au NPMs rank next in terms of enhancement efficacy, whereas other configurations of homometallic NPMs demonstrate remarkably inferior performances. Notably, the homometallic Cu structure yields the weakest R6G signal. As illustrated by the histogram in Fig. 6(d), the Ag/Cu NPM configuration delivers the most intense peaks for nearly all extracted characteristic bands of R6G, except for that at 1360 cm$^{-1}$. Under 633 nm excitation, with 1 μM R6G, it is evident that the Ag NPM significantly outperforms the Au nanoporous structure. In contrast, the Cu/Ag and other configurations show weaker enhancements, with the homometallic Cu structure yielding the lowest overall enhancement. These observations are further supported by the extracted intensities of the five characteristic Raman bands of R6G in Fig. 6(e), which clearly show that the homometallic Ag NPM produces signals approximately 2.5 times stronger than those obtained from the homometallic Au substrate. These results are consistent with the simulation data presented in Fig. 2.

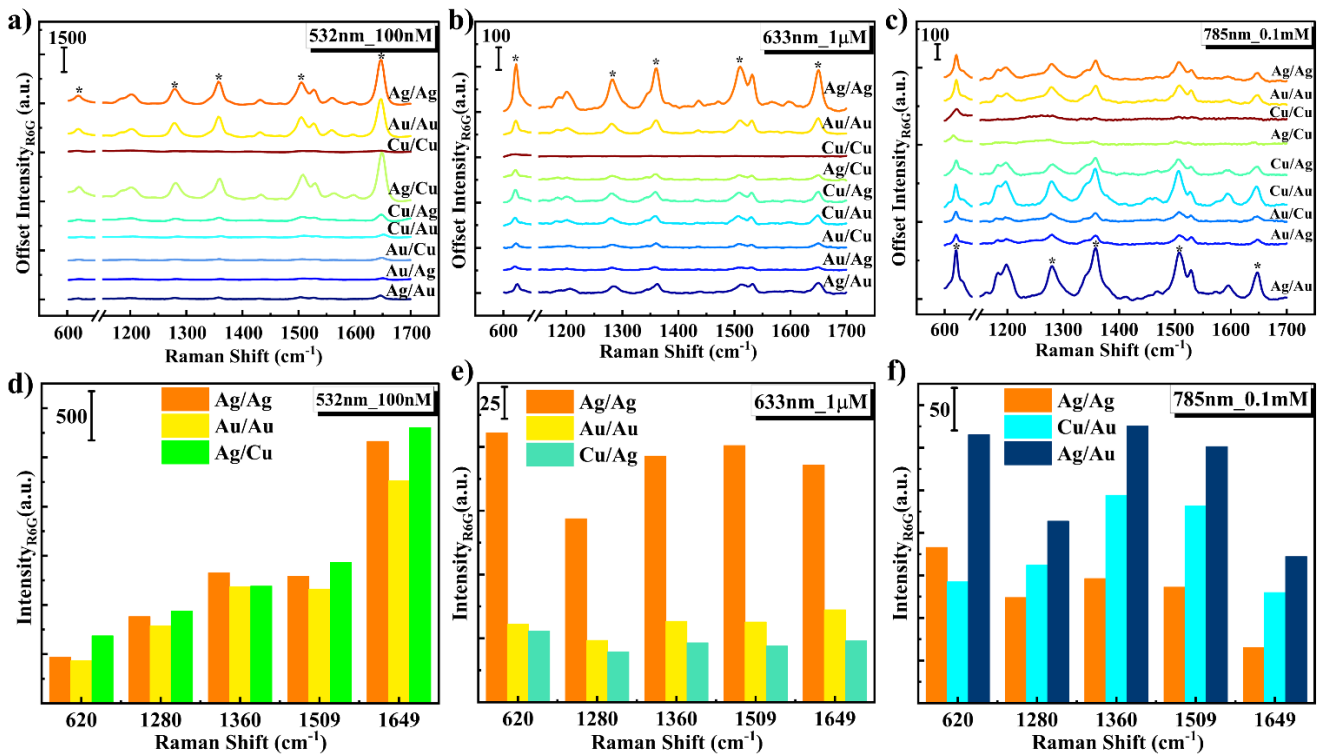

*Figure 6.* SERS spectra of R6G on various substrates under excitation wavelengths of (a) 532nm, (b) 633nm and (c) 785nm; (d-f): Comparison of the five characteristic R6G band intensities for the top three NPMs configurations identified in Figures 6 (a-c), respectively.

Under 785 nm excitation with 100 μM R6G, a different trend emerges: the Ag/Au structure (with Ag beneath Au) exhibits the strongest R6G signal enhancement, despite its negligible performance at shorter wavelengths. This result highlights the significant influence of the excitation wavelength on the SERS efficiency across different materials. Conversely, under 785 nm laser excitation, the enhancement from the homometallic Ag structure is even weaker than that of the Cu/Au structure, which appears inconsistent with the simulation results in Fig. 2. We attribute this discrepancy to the potential oxidation of the exposed Ag nanostructured top layer, as there was a significant time gap (more than one month) between substrate fabrication and measurement. In contrast, the Ag/Au structure, protected by the chemically inert Au top layer, likely maintained its plasmonic properties. To verify this hypothesis, a supplementary experiment was conducted using NPM substrates within two weeks from fabrication (see Supplementary Information), showing that the homometallic Ag structure demonstrates significantly stronger enhancement than Ag/Au, which is consistent with the simulations in Fig. 2.

To further assess SERS performance, based on the above discussion, we evaluated the signal uniformity, limit of detection (LOD), and concentration dependence of Ag/Cu, homometallic Ag (bi-layer Ag/Ag), and Ag/Au NPMs under 532 nm, 633 nm, and 785 nm excitation, respectively. Results are summarized in Figure 7. Fig. 7 (a-c) reports the averaged SERS spectra of R6G at varying concentrations measured under consistent experimental conditions. The lowest detectable concentrations of R6G were determined to be as low as 10 nM, 0.1 μM, and 10 μM for 532, 633, and 785 nm excitation wavelengths, respectively. Concentration calibration curves are also shown in Fig. 7(d-f) left panels, derived from the five characteristic SERS bands at 620, 1280, 1360, 1509, and 1649 cm$^{-1}$ in the SERS spectra shown in Fig. 7 (a-c). We found that the intensities of five characteristic SERS bands of R6G exhibit an essentially linear relationship with the logarithm of the corresponding R6G concentrations, which is consistent with the results in the literature[34]. Fig. 7 (d-f) Right panels displayed the intensity distributions of these five SERS bands across multiple randomly selected spectra on different nanoporous substrates. For each R6G characteristic peak, the relative standard deviation (RSD) of the intensities across the spectra was calculated. Subsequently, the average RSD (mean RSD) for the five peaks was determined, yielding values of 9.98%, 7.76%, and 15.53% for Ag/Cu, Ag, and Ag/Au substrates, respectively. These mean

RSD values demonstrate excellent reproducibility and uniformity of the SERS response across the different nanoporous substrates.

*Figure 7:* (a-c) Concentration-dependent SERS spectra of R6G on Ag/Cu, Ag/Ag and Ag/Au NPMs under 532, 633 and 785nm, respectively. (d-f) Left panels: Calibration curves for the five characteristic bands (620, 1280, 1360, 1509 and 1649 cm$^{-1}$) derived from the corresponding SERS spectra shown in (a-c); Right panels: Intensity distribution of these bands from randomly selected spectra for 1μM, 10 μM and 1 mM R6G on the respective substrates.

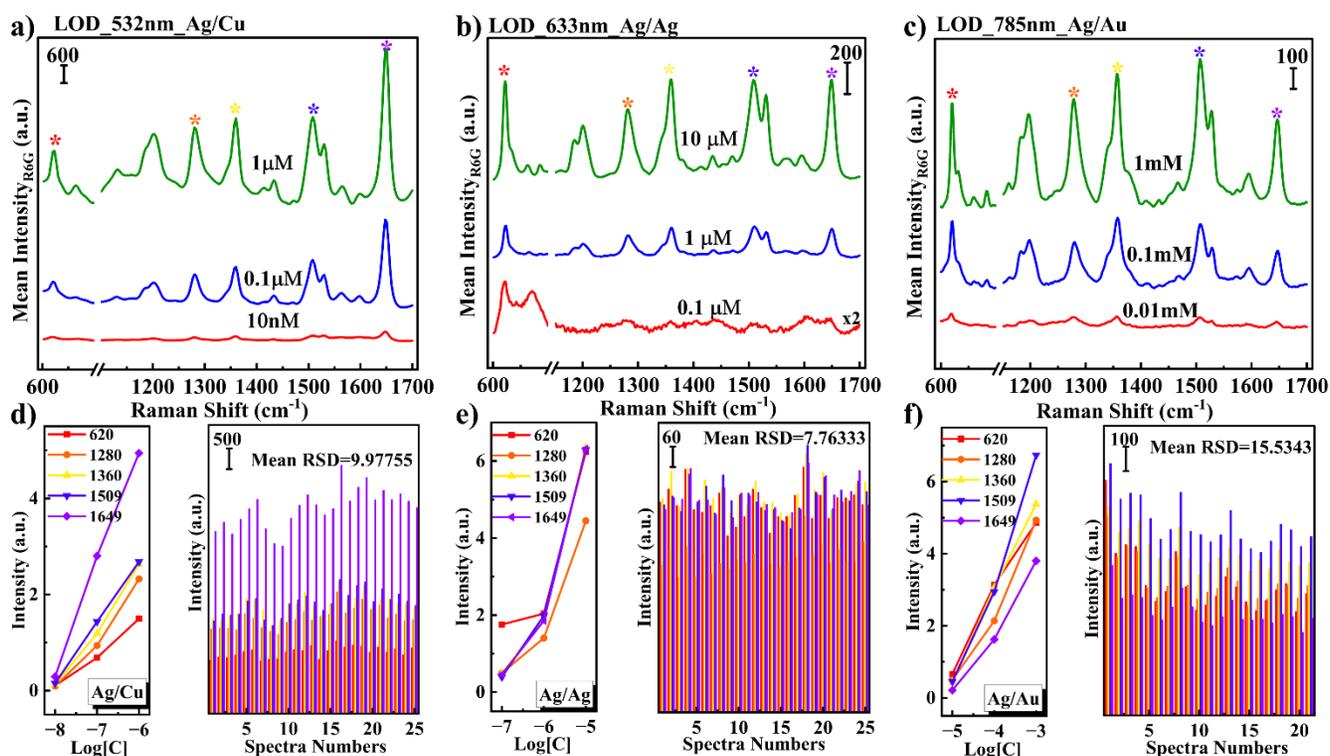

The long-term stability of the fabricated porous substrates was evaluated by measuring the SERS spectra of R6G on films stored for several months (Fig. S2, Supporting Information). Furthermore, the mechanical durability was assessed by testing the SERS enhancement after ultrasonic washing. The substrates maintained significant SERS activity even after 40 minutes of ultrasonication, demonstrating robust performance under harsh conditions (Fig. S3, Supporting Information). This combination of long-term and mechanical stability makes these substrates highly suitable for practical applications, such as photocatalytic or biochemical testing.

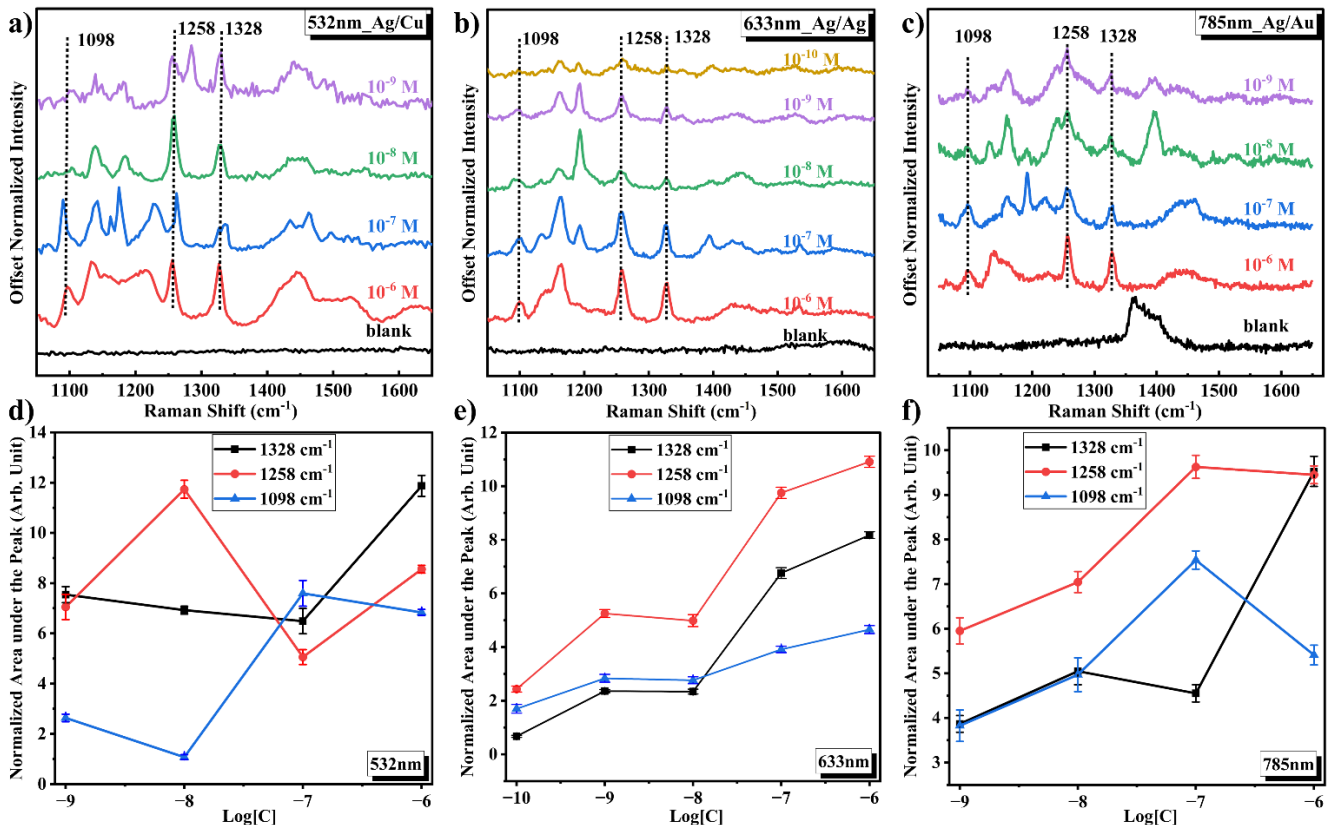

***Figure 8:*** *(a-c) Normalized SERS spectra of ADAMTS3 at various concentrations on Ag/Cu, Ag/Ag, and Ag/Au NPMs under 532, 633, and 785nm excitation, respectively. (d-f) Fit-integrated areas under the corresponding characteristics bands marked with dot lines in (a-c).*

The translation of SERS substrates from conceptual fabrication to practical application is a critical step in biosensing, particularly for the early screening and diagnosis of cancers and related biomarkers[35–37]. The ability to detect low-concentration biomarkers with high specificity and sensitivity provides a crucial, time-sensitive advantage for early diagnosis and subsequent therapeutic intervention. In a recent work, we reported a DNA origami-based SERS platform for the sensitive detection of ADAMTS3 protein, a promising biomarker associated with hepatocellular carcinoma (HCC)[38]. Here, we build upon our previous findings by deploying the novel bi-layer metal nanoporous films as a superior SERS substrate to re-investigate the detection of the same ADAMTS3 protein. As discussed above, these substrates offer large, uniform active area and long-term stability, therefore they can be interesting for applied SERS biosensing.

Figure 8 presents the analytical performance of the bi-layer NPM SERS substrates for the detection of ADAMTS3. Fig. 8 (a-c) displays the normalized mean SERS spectra of ADAMTS3 protein at various concentrations, acquired using 532, 633, and 785 nm laser excitation wavelengths, respectively. Each set of spectra corresponds to the optimally designed bi-layer configuration (e.g. Ag/Cu, homometallic Ag/Ag, and Ag/Au) for the respective laser wavelength, as detailed in the previous sections. The fitted areas under the characteristic SERS bands of ADAMTS3 protein (marked by dotted lines in a-c) at 1098, 1258, and 1328 $cm^{-1}$ were plotted in Fig. 8(d-f) as a function of logarithmic concentration with error bars. The peak fitting was performed using a Gaussian function, consistent with our previous methodology[39].

As shown in the figure, the characteristic SERS peaks of the ADAMTS3 protein remain clearly distinguishable at a concentration of $10^{-9}$ M using the three optimal enhancement substrates with the magnification of Ag/Cu, Ag/Ag, and Ag/Au under 532, 633, and 785 nm excitation, respectively. Notably, with 633 nm excitation, these characteristic peaks are still detectable even at a lower concentration of $10^{-10}$ M, representing an enhancement factor comparable to that achieved in our recent work using an optimized Au-dimer/DNA origami as a SERS platform[38]. Furthermore, the fitted areas of the characteristic peaks at 1098, 1258, and 1328 $cm^{-1}$ were plotted against the logarithmic concentration. Under 633 nm excitation, the data points exhibit smaller error bars compared to those at 532 and 785 nm, and follow an approximately linear relationship. This indicates that the peak areas increase nearly linearly with increasing protein concentration, which is consistent with our previous findings[38].

In contrast, this linear relationship is not always observed under 532 nm and 785 nm excitation. The fitted areas at 1098 and 1258 $cm^{-1}$ exhibit an approximate linear trend ranging from $10^{-7}$ to $10^{-9}$ M, but with a notable deviation occurring at $10^{-6}$ M at 785nm excitation, and all these fitted areas are associated with relatively large error bars. Under 532nm laser excitation, the relationship between the fitted area and logarithmic concentration deviates entirely from linearity for all three characteristic peaks, accompanied by significantly larger error bars compared to other wavelengths. This pronounced nonlinearity and data scatter can be attributed to the wavelength-specific electronic interaction between the protein and the corresponding bi-metallic substrate. As highlighted in the introduction section, the adsorption behavior of molecules with polar or partially charged groups—such as those present in the ADAMTS3 protein—is highly sensitive to the electronic state of the alloy surface. We hypothesize that under 532 nm excitation, the specific bi-layer configuration (e.g., Ag/Cu) exhibits an electronic structure that leads to non-uniform or saturable adsorption of protein molecules across the concentration series. This results in inconsistent SERS responses and larger variability in measured intensities, thereby disrupting the linear correlation and increasing observational uncertainties.

**Conclusions**

In summary, we conducted a comprehensive investigation of novel plasmonic platforms composed of stacked layers of noble plasmonic metals (NPMs). By preparing multiple configurations (Au/Ag, Ag/Au, Au/Cu, Cu/Au, Ag/Cu, and Cu/Ag), we were able to systematically explore how the choice of metals and their stacking order influence metal–metal interactions. Extensive numerical simulations, carried out using both finite element methods (FEM, COMSOL) and fully atomistic models, provided valuable insights into the optical properties of these nanostructures under excitation at different energies within the visible range. These models proved effective in describing the behavior of porous systems, highlighting their potential not only for enhanced spectroscopy but also for a variety of additional applications. To experimentally validate the platforms, we performed SERS measurements on both a standard probe molecule (R6G) and a biologically relevant target (the ADAMTS3 biomarker). These experiments demonstrated the strong performance of the bilayer systems as SERS substrates and offered initial evidence of how bimetallic interactions may facilitate interlayer interactions at the metal–metal interface. Altogether, this work provides a promising first step toward the study of complex multi-materials plasmonic platforms for SERS, combining the advantages of reproducibility, simplicity, and low fabrication cost.

**Author contribution**

YZ performed the Raman experiments and analyzed the data, AS and TG performed the numerical simulations, AS, AA, LB, NM and IM supported the numerical simulations and results discussion, CW and HJ supported the experimental measurements, AD, GL, SW and RK supported the samples fabrication, ZJ, ZZ, and DG supervised the work. YZ and AS equally contributed to the present work. All authors contributed to the writing of the manuscript.

**Supporting Information**

Supporting Information is available free of charge at: ……..

**Acknowledgment**

The authors thank the National Natural Science Foundation of China (No.22202167), the National Key Research and Development Project of China (No. 2023YFF0613603), and the HORIZON-MSCADN-2022: DYNAMO, grant Agreement 101072818. NM and BJ acknowledge the 'Excellence by Choice' Programme at Umeà University funded by Kempestielserna (grant no. JCK-2130.3). The authors thank the Clean Room Facility of IIT.

# 1. Additional SERS Experiments

- *R6G test under 785nm excitation with new set of substrates*

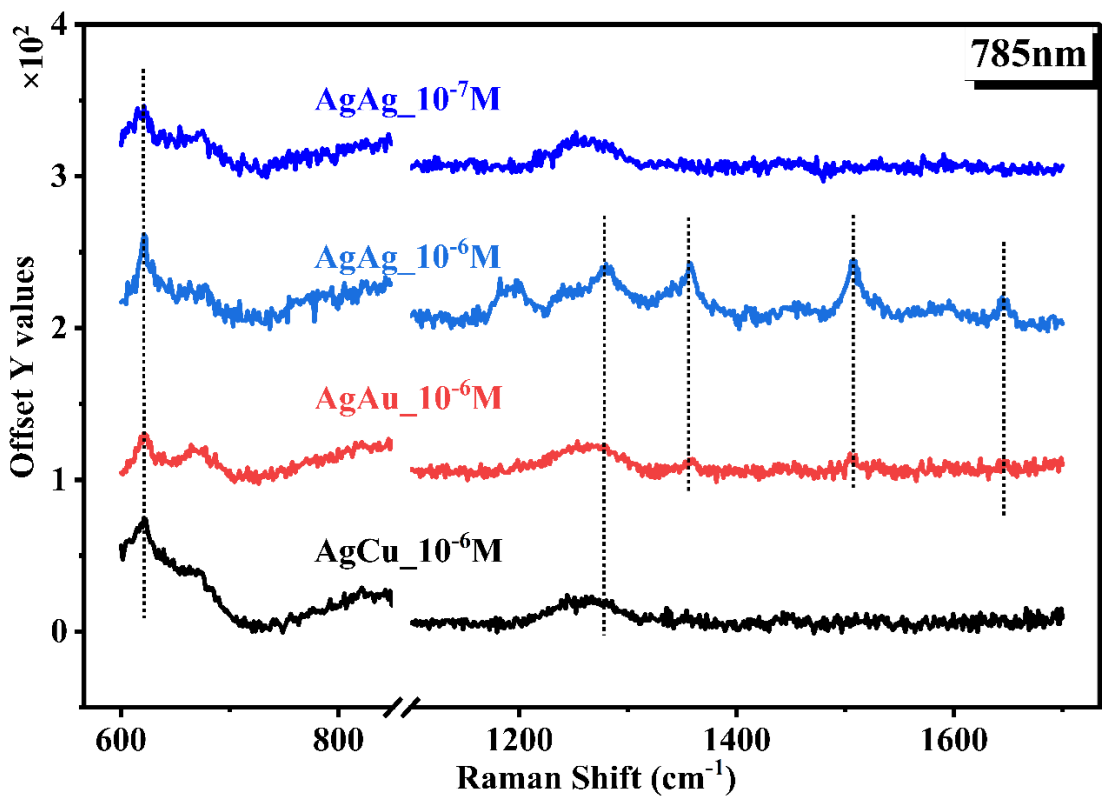

**Fig. S1.** SERS spectra of R6G at different concentrations obtained from NPMs with varying configuration (Ag/Cu, Ag/Au and Ag/Ag) under 785nm excitation.

The SERS performance of different bi-layer nanoporous substrates (Ag/Ag, Ag/Au, Ag/Cu) was re-evaluated using R6G under 785 nm excitation, as shown in *Fig. S1*. A direct comparison reveals that the Ag/Ag substrate possesses the highest enhancement capability, which is different from the previous results in *Figure 6(c)*. It successfully detected R6G at a concentration of 0.1 μM, whereas under identical conditions, the Ag/Au and Ag/Cu substrates only yielded measurable signals at a higher concentration of 1 μM.

This result clearly indicates that the Ag/Ag bi-layer structure offers superior SERS sensitivity over the other two configurations at 785nm excitation.

- *Long-term stability of the substrates*

After initial characterization, we re-evaluated the SERS performance of 100nM R6G under 532 nm excitation on various bi-layer nanoporous substrates following a two-month storage period under ambient conditions. The four substrates exhibiting the highest enhancement were selected for further analysis, as shown in *Fig. S2*. Both the bi-layer Ag and bi-layer Au substrates maintained the strongest SERS enhancement after storage. In contrast, the Au/Cu and Cu/Au configurations also demonstrated considerable enhancement following the two-month period. Notably, the Ag/Cu substrate, which initially ranked third in enhancement intensity in *Figure 6(a)* and *Figure 6(d)*, showed a significant reduction in SERS performance after storage. We attribute these observations to the protective role of gold against oxidation, which helps preserve the nano-structural integrity of the underlying porous film. In comparison, silver and copper are more susceptible to oxidation over time, leading to diminished SERS activity in Ag/Cu bilayers after prolonged exposure to air.

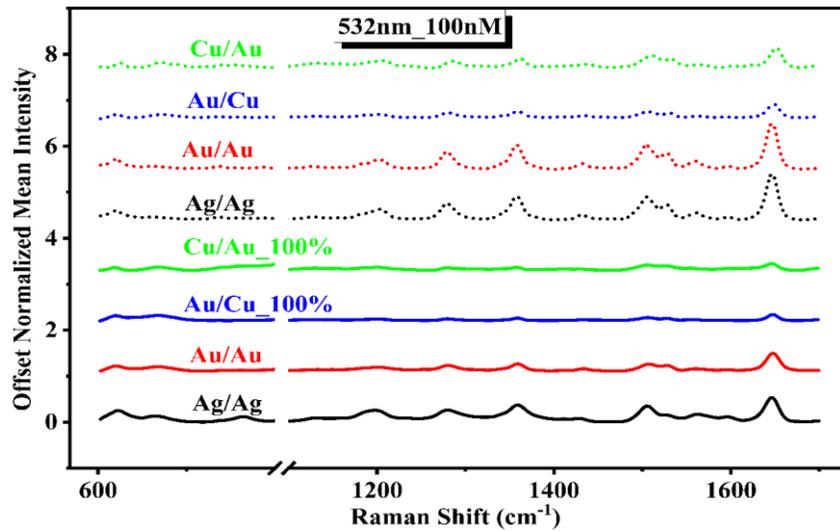

**Fig. S2.** Normalized SERS spectra of 100nM R6G acquired from the selected bilayer nanoporous substrates, before (short dot line) and after (solid line) two months storage period. Spectra were collected under 532nm excitation with all measurements using a 10% laser filter except for the post-storage Au/Cu and Cu/Au samples, for which a 100% laser power was employed. All spectra are normalized to [0,1] range for comparative purposes.

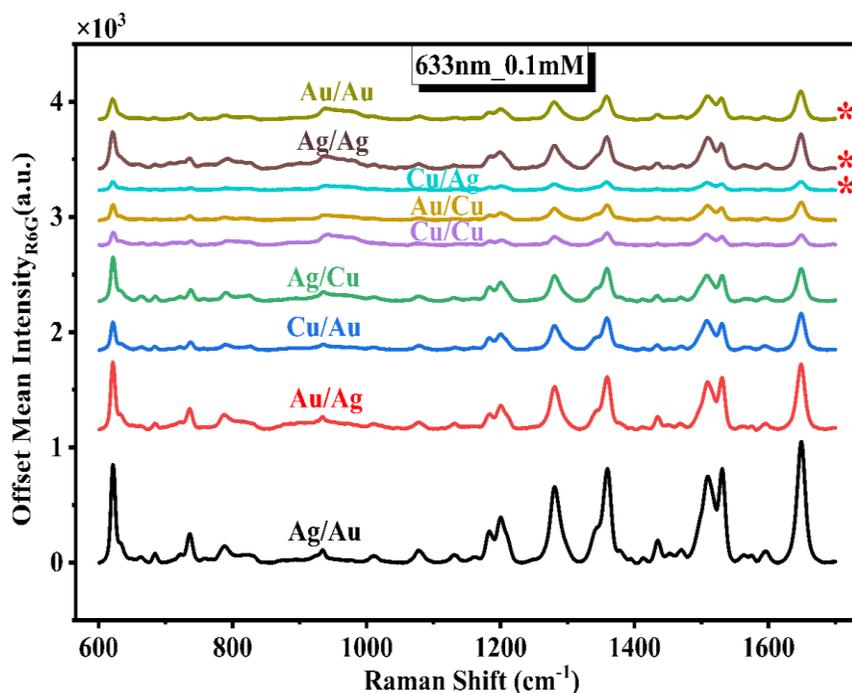

**Fig. S3** Evolution of SERS performance after storage. Mean SERS spectra of 0.1mM R6G on bi-layer nanoporous substrates with different configurations under 633nm excitation (5% laser filter), acquired after a two-month storage period.

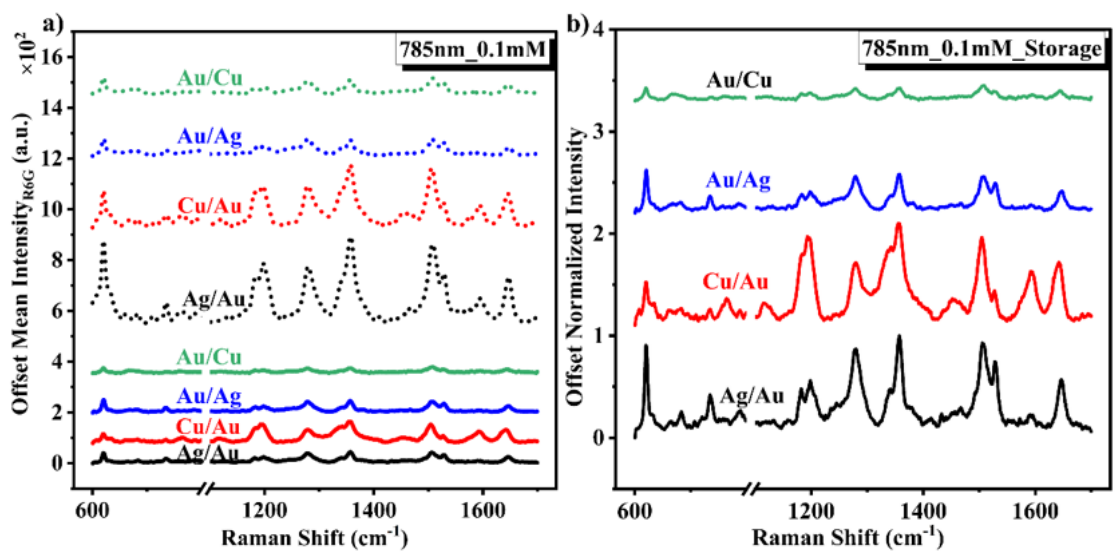

**Fig. S4** Evolution of SERS performance after storage under 785nm laser excitation. a) Mean SERS spectra of 0.1mM R6G obtained initially (dotted line) and after storage (solid line); b) Corresponding normalized post-storage spectra.

Having established the variation in initial SERS enhancement across substrates in *Figure 6(b)*, we proceeded to evaluate its subsequent evolution of the same substrates following a two-month storage period under 633nm excitation. As shown in *Fig. S3*, Ag/Au bilayer emerges as the most effective substrate, exhibiting the highest SERS intensity, followed by Au/Ag configuration. In contrast, the initially superior configurations (Ag/Ag, Au/Au, and Cu/Ag, marked by asterisks) displayed a considerable decline after storage. This marked contrast underscores the significant role of structural stability and oxidation resistance in maintaining SERS activity over time.

The results obtained under 785 nm excitation differed from those under 532 nm and 633 nm, as shown in *Fig. S4*. For comparison, we evaluated the top four enhanced SERS configurations—Ag/Au, Cu/Au, Au/Ag, and Au/Cu—before and after a two-month storage period. As illustrated in *Fig. S4(a)*, the SERS intensity of R6G decreased in all configurations between the initial and post-storage measurements. However, configurations with Ag/Au and Cu/Au consistently exhibited stronger enhancement than the others, both initially and after storage. In *Fig. S4(b)*, after normalization of the post-storage spectra, it can be observed that Ag/Au and Cu/Au configurations maintained superior enhancement performance before and after storage. In contrast, the double-layer silver porous structure, which initially ranked third in enhancement (as shown in the main text *Figure 6(c)*), exhibited weaker signals than both Au/Ag and Au/Cu after two months of storage.

- *Influence of washing to the substrates*

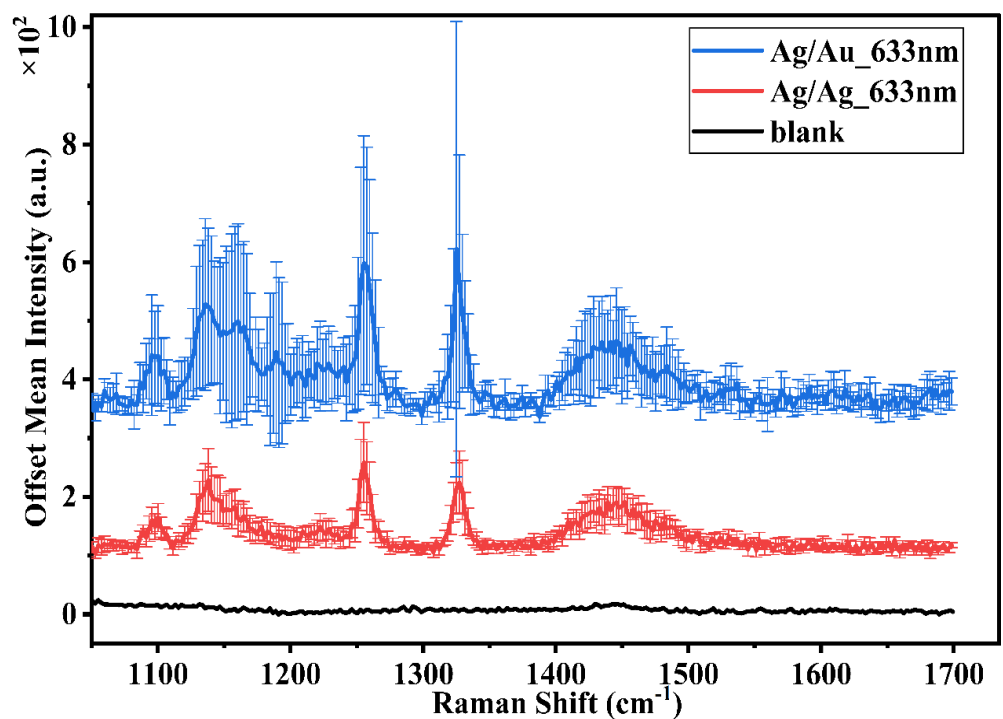

**Fig. S5** Evaluation of the washing effect on substrate stability. Mean SERS spectra of the background signal and $10^{-6}$ M ADAMTS3 were acquired under 633 nm excitation after 40 minutes of sequential ultrasonication in ethanol and ultrapure water. The results for both Ag/Au and Ag/Ag porous substrates are shown, with error bars representing the standard deviation across multiple measurements.

## 2. Additional data calculations

**Figure S6.** (a), (b), (c), (d)

Ag/Ag
Ag/Au
Au/Ag
Au/Au
NPM

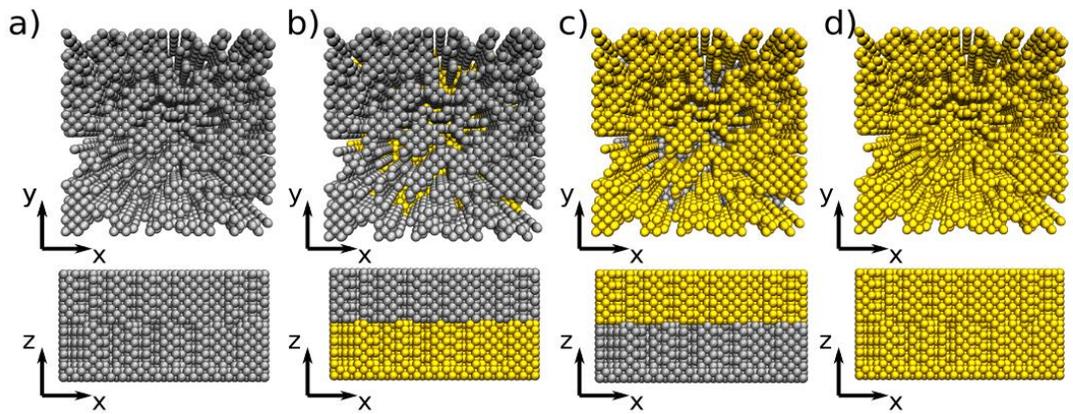

structures exploited in ωFQFμ calculations.

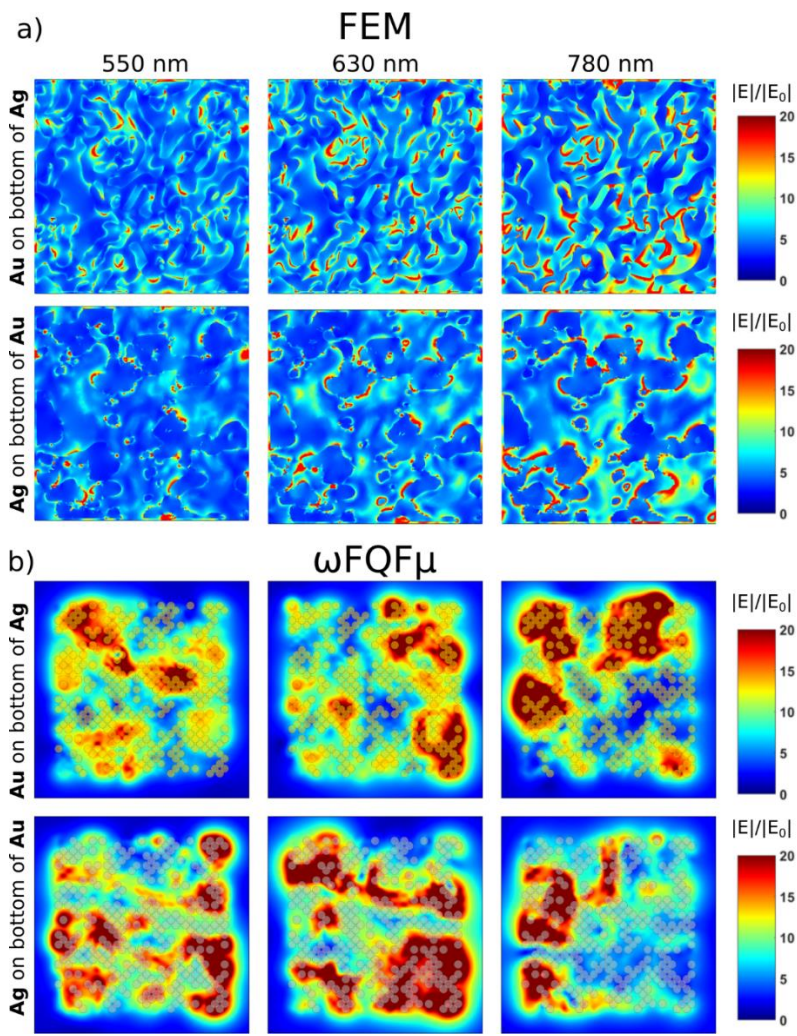

**Figure S7.** The enhanced EM field distribution of Au and Ag bottom layers for Au/Ag (top) and Ag/Au (bottom) NPMs calculated at the FEM (a) and ωFQFμ (b) levels of theory.

## 3. Ellipsometric characterization

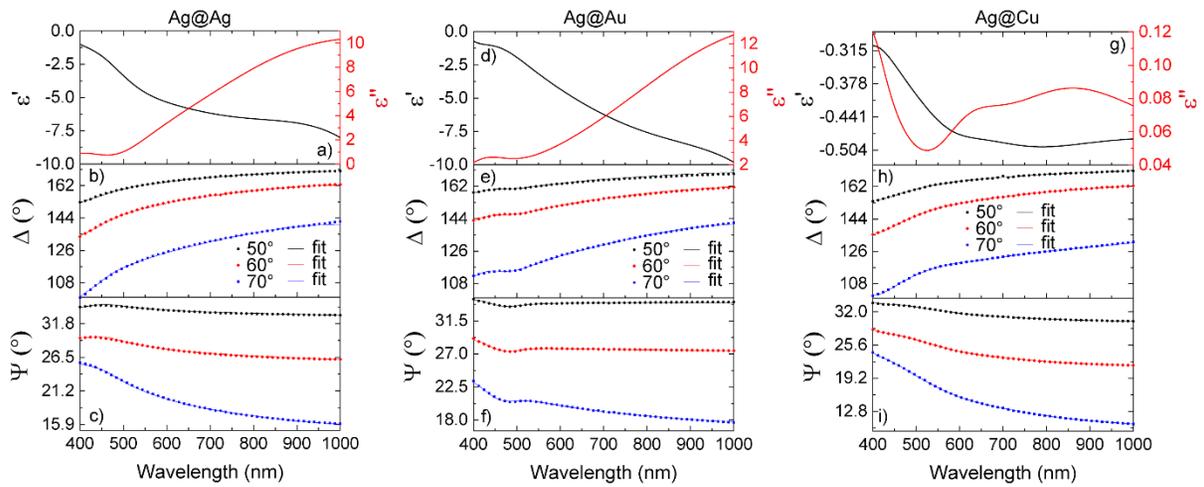

**Figure S8:** Real (black) and imaginary (red) parts of the dielectric permittivity for Ag/Ag (a), Ag/Au (d), and Ag/Cu (g) nanoporous bilayers, obtained by fitting ellipsometric data using a combination of Lorentzian and Gaussian oscillators. Corresponding experimental (dots) and fitted (solid lines) ellipsometric angles Ψ and Δ are shown in panels (b-c), (e-f), and (h-i), respectively. The oscillator parameters are reported in Table S1.

The dielectric functions of the three nanoporous bilayers, Ag/Ag, Ag/Au, and Ag/Cu, shown in Figure S8.a,d,g, respectively, were modeled using a set of Lorentzian and Gaussian oscillators to capture the key features of their optical response. For Ag/Ag (Figure S8a), the response is dominated by a single Lorentzian centered at 1.297 eV, accounting for the low-energy plasmonic resonance typical of nanostructured silver. A Gaussian oscillator with negative amplitude was included at 2.84 eV to compensate for the overestimated $\varepsilon''$ background induced by the Lorentzian tail, ensuring better agreement with the experimental data in the near-UV region. In the case of Ag/Au (Figure S8d), a broader Lorentzian ($En_0 = 1.20$ eV, $Br = 1.44$ eV) and two Gaussians were required to account for the increased damping and the presence of interband transitions characteristic of gold. The higher complexity of this system reflects the hybridized electronic structure of the alloy. Finally, the Ag/Cu (Figure S8g) sample shows a markedly suppressed optical response, with three weak oscillators spread across the visible range. The low oscillator strengths and absence of sharp features are consistent with stronger damping and reduced plasmonic activity due to the presence of copper. In all cases, the fitted dielectric functions reproduce the experimental ellipsometric parameters $\Psi$ and $\Delta$, shown in Figures S8b,c, S8e,f, and S8h,i with high fidelity across the measured spectral range.

**Table S1:** Parameters of the Lorentzian and Gaussian oscillators used to model the dielectric functions of the Ag/Ag, Ag/Au, and Ag/Cu bilayers. $En_0$: central energy (eV), Br: broadening (eV), $A_0$: amplitude (dimensionless).

| Sample | Oscillator Type | $A_0$ | $En_0$ (eV) | Br (eV) |
|---|---|---|---|---|
| Ag/Ag | Lorentzian | 10.01 | 1.297 | 0.200 |
|  | Gaussian | −1.40 | 2.840 | 1.023 |
| Ag/Au | Lorentzian | 13.17 | 1.200 | 1.440 |
|  | Gaussian | 1.07 | 2.880 | 0.644 |
|  | Gaussian | 2.98 | 4.580 | 1.745 |
| Ag/Cu | Lorentzian | 0.115 | 3.210 | 0.960 |
|  | Gaussian | 0.021 | 1.970 | 0.440 |